# Financing Entrepreneurship and Innovation in China

Lin William Cong, Charles M. C. Lee, Yuanyu Qu, and Tao Shen[*]

March 7, 2019

## Abstract

This study reports on the current state-of-affairs in the funding of entrepreneurship and innovations in China and provides a broad survey of academic findings on the subject. We also discuss the implications of these findings for public policies governing the Chinese financial system, particularly regulations governing the initial public offering (IPO) process. We also identify and discuss promising areas for future research.

[*] Cong (will.cong@chicagobooth.edu) is at the University of Chicago Booth School of Business, Chicago, IL; Lee (clee8@stanford.edu) is at the Graduate School of Business, Stanford University, Stanford, CA; Qu (quyuanyu0620@163.com) is at the School of Banking and Finance, University of International Business and Economics, Beijing, China; and, Shen (shentao@sem.tsinghua.edu.cn) is at the School of Economics and Management, Tsinghua University, Beijing, China. Tao Shen thanks the National Natural Science Foundation of China (grant ID 71603147) for financial support. Likewise, Will Cong thanks the Initiative on Global Markets and the Polsky Center at the University of Chicago Booth School of Business for research funding. We appreciate helpful discussions with Richard Lim of GSR Ventures and Pooja Malik of Nipun Capital. All errors are our own.

## 1. Introduction

Innovation and entrepreneurship rank highly on the strategic agenda of most countries today. As global competition intensifies, most national policymakers now recognize the central importance of technological advancements to long-term economic growth and societal prosperity (Abramowitz, 1956; Solow, 1957). Research shows younger firms contribute disproportionally to job creation.[1] Young firms are also more likely to experiment with disruptive technologies and business models that lead to positive knowledge spillovers (Bloom et al., 2013; Kogan et al., 2017). The cultivation and development of dynamic young firms is especially important to emerging economies, where new entrants with transformative business models can take advantage of the rapidly changing landscape in mobile-commerce and web-based technologies.[2]

No country is playing a greater role in the redrawing of the global innovation map than China. The 2018 edition of the Global Innovation Index (GII) ranks China as 17th among 126 countries by total innovation score – the highest score received by any country not in the high-income category.[3] In the past three years (2015-2017), total venture capital and private equity (VCPE) funds invested in China-based start-ups reached US$403.6 billion, making China second only to the United States as a destination for the deployment of VCPE funds.[4] In March 2018, China's Ministry of Science and Technology issued a report listing 164 Chinese "unicorns" (privately-owned firms worth more than US$1 billion each), with a combined estimated worth in excess of US$628 billion.[5]  For comparison, recent figures show 132 U.S.-based unicorns as of the end of 2017, valued at around US$700 billion.  By any of these measures, China is already a central hub of global innovation, particularly in high-tech industries.  Yet while much of this innovation is taking place through entrepreneurial ventures, little is known about how these initiatives are being financed, and to what extent financial constraints are still binding on Chinese entrepreneurs.

---

[1] Economic Co-operation and Development (OECD, 2016) statistics show that across the 21 economies studied, firms 5 years old or younger account for only 21% of total employment but are responsible for 47% of the job creation.

[2] Mobile phone adoption is disproportionally important to emerging economies. Between 2014 and 2020, an additional 1.1 billion individuals will acquire a mobile phone for the first time. At current rates of adoption, China and India will soon each have more internet users than the entire population of the United States and Western Europe combined. [Source: the OECD Science, Technology and Innovation Outlook 2016 Highlights]

[3] The GII composite score is a broad-based measure of country-level innovation, computed using 79 indicators that span both innovation related input and outputs. This annual report is jointly produced by Cornell University, INSEAD, and the World Intellectual Property Organization (WIPO). The upper-echelons of the GII ranks are dominated by high-income countries (as measured by per capital GDP). Among the top 30 countries ranked by total GII score, China alone is an upper-middle income country. Only two other upper-middle income countries (Malaysia and Bulgaria) ranked in the top 40. For a detailed discussion of the conceptual framework behind the GII, see Casanova, Cornelius, and Dutton (2018).

[4] The distinction between venture capital (VC) and private equity (PE) activities is blurred in China. We therefore refer to both together as VCPE investing. In Section 3 we present more details on total VCPE investments.

In this study, we provide an overview of the current state-of-affairs in the financing of private innovations in China.  While country-level innovation can take many forms, our focus is on the funding of business start-ups and entrepreneurial ventures.  The funding needs of a start-up business will vary over its entrepreneurial life cycle.   For example, Figure 1 provides a graphic representation of the different funding sources commonly available to a business enterprise at each stage of its life cycle.  Using this figure as an organizing framework, we survey, and offer an evaluative commentary on, the funding sources that are available to a Chinese firm during each stage of its life cycle: early/seed stage; medium/growth stage; and late/expansion stage.

Our study has four specific objectives: (a) to present an economic framework for evaluating the central challenges associated with the financing of entrepreneurial ventures in China, (b) to evaluate the relative size and importance of the channels through which private initiatives for innovation in China are currently being funded, (c) to survey the academic evidence on potential financing constraints currently facing private initiatives in innovation, and (d) to discuss public policy implications that may arise from these findings, as well as to outline the type of future research that may best inform Chinese policy makers.

We begin in Section 2 with a review of the central economic themes in entrepreneurial finance. This analysis identifies three key forces that shape the economics of venture investing: (a) the non-rival nature of the output, (b) the high uncertainty and payoff skewness associated with the process, and (c) the large potential for agency conflicts between entrepreneur and financier, arising from information asymmetry and moral hazard issues. These three forces are at the root of the most vexing challenges in financing for innovation. Indeed, many features of the modern-day VCPE industry, as well government policies on patent protection, subsidies, and tax incentives, can be understood as efforts to mitigate the negative externalities associated with these economic problems.

---

[5] *The 2017 China Unicorn Enterprise Development Report*, jointly released March 20, 2018, by the Torch High Technology Industry Development Center of the Ministry of Science and Technology and the Greatwall Strategy Consultants in Beijing.  For a complete list of these companies, see http://westdollar.com/sbdm/finance/news/1345,20180323847480686.html.

We proceed in Section 3 with a review of the channels through which external funding now reach entrepreneurs in China. We show that VCPE funding has increased exponentially in recent years. Most of this funding is domestic, but a sizeable amount comes from overseas. Furthermore, government entities and state-owned enterprises (SOEs) are also significant direct investors in many start-ups. Our evidence suggests that "mid-stage" financing, covering the expansion and growth stages of a firm's life cycle (see Figure 1), may not be a major problem for Chinese entrepreneurs. On the other hand, compared to start-ups in the United States, early-stage (seed) funding may still be more difficult to secure in China. More importantly, our analyses led us to focus sharply on "late-stage" financing, and the exit strategies available to Chinese entrepreneurs. In particularly, we identify a number of problems with China's antiquated initial public offering (IPO) regulations. In our view, these regulations now loom as a significant obstacle to entrepreneurship and innovation in China.

Section 4 further explores the problems engendered by China's IPO regulations. In contrast to the registration-and-disclosure system that exists in most countries, the IPO process in China is strictly regulated. Candidate firms are required to meet strict pre-specified profitability and revenue thresholds. Firms meeting these standards typically face a further waiting period, as the China Securities Regulatory Commission (CSRC) reviews, and adjudicates on, every applicant. This process is arduous, the outcome is far from certain, and there is mounting evidence that the ultimate decision is not determined solely on economic merit.[6] Perhaps most importantly, the total number of firms allowed to IPO in a given time period (the IPO "quota") is tightly controlled by government policy, often leading to large backlogs of firms awaiting review,

---

[6] Studies that suggest political connections play a role in China's IPO allocation decisions include Fan, Wong, Zhang (2007), Francis, Hasan, Sun (2009), Piotroski and Zhang (2014), Li and Zhou (2015), and Lee, Qu, and Shen (2018b).

particularly after an IPO "suspension" period.[7]

Based on our survey of the academic evidence, as well as further empirical analyses conducted in this study, we identify a list of economic problems and consequences that are directly attributable to China's current IPO regulations:

1. Long wait times and substantial outcome uncertainty for candidate firms seeking access to domestic equity markets

2. A bias against high-growth technology firms, which typically have lower profits, less developed businesses, and more intangible assets

3. Substantial underpricing of IPOs, resulting in exceptionally large initial-day returns that dwarf those seen in more developed markets

4. An exodus of high-quality, particularly high-technology, candidate firms to foreign equity markets

5. Costly reverse merger (RM) transactions in which highly-qualified, but less politically-connected private firms pay more than US$400 million each, on average, for a listed shell company

6. Virtually no delisting or retirement of failed companies from public equity markets (in fact, these failed businesses continue to propagate by levering their listing status to acquire new lines of business, thus maintaining control and circumventing the IPO process)

7. Large cross-sectional price distortions among publicly listed firms (including systemic risk associated with IPO regulations, and an enormous Size premium for the smallest listed firms, which trade more on their expected shell value than on corporate profits)

8. Listing delays in the IPO process that lead directly to a reduction in firms' innovation activity, as measured by patent quantity and quality (such effects begin during the delay period and endure for many years after listing)

9. Potentially inflated market prices for all publicly listed firms, as well as higher levels of speculative trading by domestic investors

---

[7] Between 2004 and 2016, the CSRC suspended all IPO activities on five occasions. These suspensions lasted between six and fifteen months each. The specific timing of these IPO suspensions and reboots can be found on a CSRC authorized website, http://stock.cnstock.com/stock/smk_gszbs/201701/4013651.htm. These suspensions typically lead to large review backlogs. For example, as of the end of October 2016, companies meeting China's pre-specified listing standards and awaiting CSRC processing numbered 806. For reference, the number of firms that approved for listing in the first 10 months of 2016 averaged 7.7 per month.

In Section 5, we summarize our findings, discuss policy implications, and explore potential venues for future research. We conclude that China's current IPO regulations represent a serious impediment to two important near-term goals espoused by the Chinese government: (a) To bring more high-technology firms back to mainland stock markets, and (b) to be included at a meaningful weight in international stock indices, particularly the MSCI Emerging Market Index.

Considering the problems identified above, we recommend a move toward a registration-and-disclosure system for Chinese IPOs, like those employed by most other countries. In such systems, investors monitor firm quality and market forces adjudicate firm value. Firms that receive enough support from the investment community will attain IPO status. The role of regulators is to ensure adherence to established ordinances, which are largely disclosure-centric.

We acknowledge the need to protect retail investors and minority shareholder. But this protection need not come by limiting the access of startup firms to public equity markets. Instead, the protection can be in the form of more stringent enforcement of insider trading laws, increased corporate transparency and quality of disclosure, and changes in the judicial system that would facilitate swift recourse through private litigation in the event of majority shareholder misconduct. None of these reforms would require regulators to adjudicate firms' investment value; a task that we believe is best left to markets.

## 2. The Economic Underpinnings of Entrepreneurial Finance

The financing of innovation-based projects gives rise to three important classes of economic problems that have been widely studied in the literature. In this section, we discuss these problems and provide a brief survey of key studies in each area. We also discuss how these central economic problems are related to the rise of venture capital (VC) funding, as well as the set of public policies commonly observed in this domain.

First and foremost are the challenges arising from the *non-rival nature of the output*. Innovation and experimentation are costly undertakings whose payoffs often come in the form of knowledge, which is a non-rival good (Schumpeter, 1942; Arrow, 1962). While the innovator and her funding partners bear the risks and costs associated with a project, much of the benefit derived from the undertaking can accrue to others, through knowledge spill-overs. Problems associated incomplete appropriability of investment returns is a key reason why innovations can be difficult to finance with capital from sources external to the innovator or entrepreneur. Yet rivalry and excludability of technological innovations play crucial roles in a country's endogenous growth (Romer, 1990). These problems can be partially mitigated through public policies, for example those governing intellectual property protection, government subsidies, or tax incentives.[8] Nevertheless, even with such policies in place, other factors that affect an innovator's ability to capture profits generated by an innovation can still have a significant effect on incentives for innovation in an economy.

A second set of challenges arise from the *risky nature of the process.* Innovation and experimentation are speculative activities, and the associated payoffs often involved a high level of uncertainty that is difficult to quantify ex ante. As the risk of the venture increases, investors' ability to manage these problems through purchase discounts alone becomes increasingly limited. In highly speculative ventures with imperfect information, a host of agency issues (Stiglitz and Weiss, 1981) and hold-up related problems (Ewens, Rhodes-Kropf, and Strebulaev, 2016; Hochberg, Ljungqvist, and Vissing-Jørgensen, 2013) can arise, requiring investors to revisit the terms of the agreement. Such needs have given rise to more complicated contracts in the domain of venture investing, often featuring contingency claims. By making the extent of the manager's control and access to capital contingent on performance, such claims can help mitigate agency concerns that increase with issuer risk. Of course, to the extent that such risk-sharing problems cannot be fully resolved through contingent contracts, a certain amount of under-investment in innovative activities will still result.

A third set of challenges arise from the *agency conflicts* between entrepreneurs and investors, in the form of information asymmetry (Akerlof, 1970); Leland and Pyle, 1977) and moral hazard problems (Jensen and Meckling, 1976). These agency- and information-related issues are particularly salient for young, small firms, and in developing countries such as China where the financial markets are not fully developed, and the regulatory landscape is complex. In such settings, providers of capital will need acquire a certain minimal level of expertise in the innovative activities at hand. This applies to both private investors and public entities that seek to finance the project. But of course, innovation-based projects are, by nature, more difficult to evaluate, leading potentially to sub-optimal allocations of resources.

---

[8] Romer (1986) led the literature of endogenous growth models that justify government interventions, such as R&D tax incentives, in innovation-related processes.

A large literature has been devoted to how financial intermediaries, especially venture capital (VC) firms, can help overcome the aforementioned challenges and thus facilitate innovation. Prior studies indicate VC firms can improve efficiency through the active monitoring and advice provided (Hellmann, 1998), the screening mechanisms employed (Chan, 1983), the incentives to exit (Berglof (1994)), the proper syndication of the investment (Admati and Pfleiderer, 1994; Lerner, 1994), superior investment staging prior to exit (Bergemann and Hege, 1998; Sahlman, 1990; Gompers, 1995; Tian, 2011b), or better governance once the firms become public (Lerner, 1995; Hochberg, 2011). Overall, theory and evidence to date indicate that venture capital firms can improve monitoring, and help start-ups in standardization and professionalization their businesses (Hellmann and Puri, 2002; Rajan, 2012; Cong, Howell, & Zhang, 2017), thereby improving outcomes and potentially encouraging more innovation (Kortum and Lerner, 2000; Bernstein, Giroud, and Townsend, 2016).

A related line of research examines how mechanism design can better align incentives between entrepreneurs and financier, thus mitigating agency problems. The specific mechanisms studied include security design (Cornelli and Yosha, 2003), as well as experimentation, manager compensation and style (Manso, 2011; Tian and Wang, 2011). Several studies also examine mitigation mechanisms for the hold-up problem in the venture financing setting (Ewens, Rhodes-Kropf, and Strebulaev, 2016; Hochberg, Ljungqvist, and Vissing-Jørgensen, 2013). Most recently, Azarmsa and Cong (2018) explore a combination of information design and security design.

In addition to considering the social value of innovation, it is also important to understand the economic incentives for venture investing from the perspective of financiers and entrepreneurs.

Towards this end, a large literature has developed on the risk-return tradeoffs in venture investing (Korteweg and Sorensen, 2017; Cochrane, 2005; Gompers and Lerner, 1997; Jones and Rhodes-Kropf, 2003). One branch of this literature focuses on evaluating performance persistence at the fund level (Kaplan and Schoar, 2005; Harris, Jenkinson, Kaplan, and Stucke, 2014), individual partner level (Ewens and Rhodes-Kropf, 2015), or deal level (Nanda, Samila, and Sorenson, 2018). Recent debates have centered on luck versus skill in VC performance, taking into consideration the endogenous nature of deal flows (Sørensen, 2007), and the interaction of flows with fund manager contracts (Cong and Xiao, 2018). Overall, this literature finds performance persistence in the VC industry, with managerial skill and endogenous deal flows both playing a role. However, debate continues on whether the asset class outperforms other asset classes after adjusting for risk. Relatively less explored is the value proposition of ventures from an entrepreneur's perspective, with Manso (2016) being a notable exception.

To better understand the returns to financing innovation and how financiers add value beyond capital provision, researchers have investigated what exactly venture capitalists do. A key early study is Kaplan and Strömberg (2004), which focused on contracting-related benefits. More recent studies have discussed the decision making process of the typical VC fund (Gompers et.al., 2016), as well as valuation issues that VCs often face (Gornall and Strebulaev, 2017).

One particularly important decision financiers have to make involves the exit strategy. It is well-known that younger venture firms have a tendency to rush the incubation period and thus underprice IPOs, in order to build their reputation and improve prospects for future fund-raising (this phenomenon is described in the literature as "grandstanding", see, e.g., Gompers (1995), Lee and Wahal (2004), and Grenadier and Malenko (2011)). Equally interesting is the process by which venture firms gradually transfer public shares to their investors (Gompers and Lerner, 2000). On balance, alternative exit strategies and reasons for failure are relatively understudied areas of venture investing, with Puri and Zarutskie (2012) being a notable exception.

Private VC firms are not the only source of venture capital funding for early- and mid-stage entrepreneurs. A few papers contrast VC to other alternative funding sources such as bank lending (Landier, 2003) and angel investors (Chemmanur and Chen, 2014; Kerr, Lerner, and Schoar, 2011; Prowse, 1998). Another increasingly important funding source is corporate venture capital (CVC) from more established firms.[9] CVC allows for strategic venture investing and often bring the benefits of vertical integration (Hellmann, 2002). It can also help attract and retain talent (de Bettignies and Chemla, 2008). Gompers and Lerner (2000) provides an overview of early CVCs. Government venture capital (GVC) are also established to finance social projects and encourage innovations, especially in developing economies (Zhang, 2017; Zhang and Mayes, 2018).

In sum, three key forces shape the economics of venture financing: (a) the non-rival nature of the output, (b) the high-uncertainty level associated with the process, and (c) the large potential for agency conflicts between entrepreneur and financier, arising from information asymmetry and moral hazard issues. Many existing regulations on patent protection, government subsidies, and tax incentives can be seen as efforts to mitigate negative externalities (market failures) associated with these economic forces. Likewise, the rise of the modern-day venture capital fund, with its focus on general partners (GPs) that possess deep entrepreneurial expertise, can be viewed as a natural market response to the challenges engendered by these forces.

This summary of the literature on entrepreneurial finance is brief by necessity. We refer interested readers to several surveys that discuss research, as well as institutional details, related to venture capital investing and the financing of innovations around the world (Sahlman, 1990; Metrick and Yasuda, 2010; Hall and Lerner, 2010; Da Rin, Hellmann, and Puri, 2013; Casanova, Cornelius, and Dutta, 2018).

In the context of China, the economic underpinnings of financing innovation are further complicated by two distinct features. First, Chinese innovations have traditionally centered around improvements in business models, product modifications, manufacturing processes, and creative scaling, as opposed to inventions and breakthroughs of a fundamental nature. At first blush, this type of innovation might seem less important or significant. However, as Schumpeter (1947) aptly noted, there are many categories of innovation. Innovations of the ``one to many'' variety could be as important as those of the ``zero to one'' variety (Theil and Masters, 2014). Moreover, the nature of innovations in China has continued to evolve, with many Chinese technology firms now leading the world in the fields of artificial intelligence and FinTech.

---

[9] We use the term CVC to refer to any corporate equity investment in a startup venture. In our usage, CVC includes not only corporate venture capital affiliates (e.g. Gompers and Lerner, 1998, 2000), but also financing by small business investment companies (SBICs) as well as direct investments from larger corporations (such as Tencent or Alibaba), possibly for strategic reasons. These may be minority equity investments, or majority stakes.

Second, there are some Chinese cultural, institutional, and political contexts that may alter the nature of, and solutions to, the aforementioned problems. For example, the Chinese government plays an active role in shaping the venture capital industry: (1) by regulating foreign investments, and (2) by facilitating and encouraging investments that are more congruent to local governments' agendas. We touch on some of these issues in this study. At the same time, we hasten to point out that many of these issues remain understudied, and could constitute interesting targets for future research.

## 3. Financing Entrepreneurship in China

Having identified the central challenges in financing innovation-related activities, we now provide some direct evidence on the current state-of-affairs in China. Funding sources that are available to a company vary over its entrepreneurial life cycle. Figure 1 provides a graphic representation of the different funding sources commonly available to a business enterprise at each stage of its life cycle. In the following discussion, we group the various funding sources for a Chinese firm in terms of their impact on three stages of its life cycle: early/seed stage; medium/growth stage; and late/expansion stage.

### 3.1 Early-Stage Financing

In the earliest part of its life cycle (the "Seed" stage), a startup typically relies on the entrepreneur's personal savings or funds from friends and family. In China, as in most other emerging economies, startups have limited access to bank loans. This is due in part to the fact that the typical startup does not have sufficient tangible assets to pledge as collateral. Even when such assets exist, many banks are reluctant to lend to startup businesses either because of insufficient credit information, or ill-defined legal rights with respect to secure transactions in the event of default.  Therefore, unless other external financing becomes available, many startups fail to grow and reach positive cash flow status after the entrepreneur exhausts funding from his/her personal connections.  Casanova, Cornelius, and Dutta (2018) refer to the trough in this graph – the time between idea origination and the first VC check – as the "Valley of Death."

In more advanced economies such as the United States, would-be entrepreneurs can tap a broad set of early-stage external financing sources such as incubator and accelerator programs.  Some may have to access angel investors, microcredit, or crowdfunding sources; others may be fortunate enough to receive funding from early-stage VCs or government subsidies.  In China, some peer-to-peer financing is available through a host of crowdfunding platforms, most of which have emerged in the last few years.  Nevertheless, our sense from discussions with Chinese entrepreneurs and VCPE investors is that the procurement of early-stage financing remains more challenging in China than in the United States.

According to Floship.com, a Global Cross-Border service provider, there are more than 400 crowdfunding platforms operating in China in 2016.[10]  A vast majority of the projects available on these sites are reward- (or product-) based, rather than equity-based. The peer-to-peer (P2P) lending industry in China has also expanded rapidly to involve over 5,000 platforms from 2010 to 2016, over half of which have since failed.   In terms of the total volume of funds going to entrepreneurs, these crowdsourcing sources are still dwarfed by traditional VCs.

One recent Chinese government report lists the total amount of product-based crowdfunding in the first half of 2016 as 4.1 billion RMB (over $588 million in USD).[11]  For comparison, Kickstarter reports that it has raised $3.69 billion USD in successful projects as of the time of this writing.[12]  The total raised from equity-based crowdfunding efforts is much smaller in both countries.  Angel Crunch ([www.angelcrunch.com](www.angelcrunch.com)), which bills itself as China's largest equity crowdfunding platform, touts having raised 300 million RMB (over $43 million in USD) for startup companies thus far.  These numbers show that equity-based crowdfunding remains a minor part of the financing activities of Chinese startup businesses.

---

[10] [https://www.floship.com/crowdfunding-in-china-practices-and-trends/](https://www.floship.com/crowdfunding-in-china-practices-and-trends/)

[11] [http://www.zczj.com/news/2016-08-01/content_8140.html](http://www.zczj.com/news/2016-08-01/content_8140.html)

[12] [https://www.kickstarter.com/help/stats](https://www.kickstarter.com/help/stats)

One Chinese initiative to address early-stage funding for R&D is the Innovation Fund for Small and Medium Technology-based Enterprises (the SMTE "Innofund").  This fund was established by the State Council in May 1999 with a mandate to "facilitate and encourage the innovation activities of small and medium technology-based enterprises (SMTEs) and the commercialization of this research by way of financing, trying to bring along and attract outside financing for corporate R&D investment of SMTEs." Eligible applicants are businesses with less than 500 employees, at least 30% of whom have received college-level education.  The annual R&D investment of the firm must exceed 3% of total sales, and the number employees engaged in R&D must be at least 10% of the total work force.  From 1999 to 2011, the fund provided more than 19.2 billion RMB (approximately US$3.1 billion) to 30,537 projects.  According to literature from the fund, the program has created more than 450,000 new jobs and generated 209.2billion RMB in sales, 22.5 billion RMB in tax income, and 3.4 billion RMB in exports. As of the end of 2008, 82 out of 273 publicly-listed companies on the SME Board of the Shenzhen Stock Exchange were once supported by the Innofund.

Guo et al. (2016) investigate the effects of this program on firm innovation outputs. The authors find that Innofund-backed firms generate higher innovation-related outputs, including both commercialized outputs as measured by sales and exports from new products, as well as technological outputs as measured by the number patents.  These results hold in the cross-section (when compared to non-Innofund-supported counterparts), as well as in time-series (when compared to the same firms prior to securing Innofund support).  Overall, this study suggests that government support of R&D through the fund is effective in stimulating innovation in China.

Wang et al. (2017) also study the Innofund program, and report more circumscribed findings.  Using internal administrative data on applications to the program, this study aims to: (a) identify the application features associated with higher chances of obtaining grants, and (b) parse out the causal impact of receiving a grant on firm performance.  Although the authors find that firms with observable merits are more likely to receive funding, they also show that bureaucratic intervention plays a role in the screening process. For example, firms with political connections are more likely to receive the funding after controlling for other factors. Furthermore, after controlling for ex ante characteristics of the applicant, they find no evidence of a causal relation between grant receipt and better firm performance. This result suggests while the Innofund was able to identify firms with more promising projects, the grant itself had limited causal impact on subsequent performance. In other words, the positive impact on innovation and firm performance documented in Guo et al. (2013) is likely to be driven by selection, rather than treatment, effects.

In sum, early-stage financing is a significant challenge for entrepreneurs. Companies at the idea-stage are much less likely to draw the attention of VC funds than those with working prototypes or other evidence of technological feasibility. In fact, it is possible for early-stage startups be struggling for funding, even as VC capital reaches all-time highs.[13] These concerns suggest a greater role for government-based incubator programs such as the Innofund, or some other organizations that can also provide mentorship, advice, and connections. Unfortunately, there is a paucity of research on the efficacy of early-stage funding programs. Clearly, there is an opportunity for academics to contribute to the debate going forward.

*3.2 Mid-Stage Financing*

Firms that manage to survive the early-stage will find many more dance partners in the funding arena. Mid-stage financing is the traditional domain of venture capital funds. In addition to the usual players, venture investing in China features significant government involvement. Government entities (both central and provincial) have their own VC funds, in which they act as the general partner (GP). We will refer to these entities as Government Venture Capital (GVC) funds. Many other government entities, or state-owned enterprises (SOEs), also engage directly in venture investing. Large corporations, most notably Tencents and Alibaba, have been active in corporate venture investing (CVC). And finally, in recent years small Chinese startup firms can also apply for listing on the National Equities Exchange and Quotations (NEEQ), which is similar to the over-the-counter market in the United States.[14] We provide some evidence on the scale and impact of each of these mid-stage initiatives.

---

[13] Khosravi (2018b) notes that during the first quarter of 2018, more than US$21 billion in VC funding was deployed across 1,200 deals in the US, but the money was divided among fewer companies. In fact, in the same quarter US VC funds actually reduced their investments in seed and early stage firms. The author argues a key reason for this phenomenon is that given their fee structures, most VC firms simply do not have sufficient economic incentive to invest time and capital on relatively small/early stage companies.

Table 1 reports the number of VCPE investment events in Chinese startup firms during the 2006 to 2017 period, and the dollar value of these investments. This information is extracted from PEdata. In this database, each bilateral investor-investee transaction is defined as a separate investment event. All Chinese startup firms in the database are included. We define Chinese firms as those whose headquarters are in China; similarly, Chinese (Foreign) VCPE firms are those headquartered in China (overseas). All monetary amounts are in billions of U.S. dollars. Values originally reported in RMB or other foreign currencies were converted to US dollars using exchange rates at the end of each year. Investment values are reported as of each financing-round. When several funds participated in the same round but PEdata did not parse out the amount for each fund, we assume the funds contributed equally. We report total annual investments, as well as the cumulative amount invested over the entire 12-year period.
This table shows that venture investing in China has grown quickly in recent years. Of the total US$ 737.23 billion invested by VCPE entities, US$397.35 billion (53.9%) took place in the most recent three years. Foreign participation over this period totaled US$98.22 billion, but most of the funding (US$639.01 billion, or 86.7%) is coming from Chinese sources. In 2017, VCPE investments totaled US$151.69 billion, which is 1.75% of the total market capitalization of the listed stocks on the Shanghai and Shenzhen stock exchanges as of the end of the year.

To better understand funding sources used by China's leading private technology firms, Table 2 presents a list of the top 20 privately-owned Chinese tech firms ("Unicorns") as of the end of 2017. The list is derived from China's Ministry of Science and Technology report titled "2017 China Unicorn Enterprise Development Report." For each firm, this report provided the estimated market value (in US$B), the year of founding (Year), its industry section (Sector). We

---

[14] The NEEQ facilitates trading in the shares of a company that is not listed on either the SZSE or the SHSE. Note that listing on the NEEQ is not a financing event. We discuss the current level of trading activity in the NEEQ later in this section.

also secured additional information from PEdata, including whether each firm received funding from venture capital and private equity investors (VCPE), whether it has a state-owned enterprise as an investor (SOE), and if so, the type of SOE.  Specifically, CIC denotes the China Investment Corporation (China's sovereign wealth fund); Corp denotes a government owned corporation, Pension denotes China's National Social Security Fund, Local denotes a local government fund; and GVC denotes a government venture capital or private equity fund.

All 20 "unicorns" are valued at US$5 billion or more, and their collective valuation is more than US$400 billion.  The most heavily represented industry sectors are: Internet Finance (7), New energy automobiles (3), E-commerce (3), Internet Services (2), and Healthcare (2).  Note that all 20 have received some VCPE funding, and 12 out of 20 (60%) also received equity funding from a government entity.  China's sovereign wealth fund (CIC) and other SOEs (Corp) feature prominently as investors.  Local governments, government venture capital funds (GVC), and government-owned pension funds, all participated.  Clearly, government entities are active participants in venture investing in China.

Are government-backed ventures more likely to succeed?  Cao et al. (2013) examine this question, and find that the rate of success varies depending on whether the investing entity is managed wholly, or only partially, by a government-owned general partner (GP).  Companies backed by a partially government-owned GP are significantly more likely to achieve an IPO, and are more capable of exiting even in market downturns. However, backing by a wholly government-owned GP seems to detract from venture success. Also, collectively, government-backed VCs deliver lower returns as measured by the fund's average exit multiple.  These results suggest that some level of government ownership may help investees navigate China's heavily-regulated IPO process.  Note that while the inferior overall performance of GVCs may reflect a lack of investing expertise, it is also consistent with profit maximization not being the sole goal of GVC funds.[15]

---

[15] Relatedly, Zhang and Mayers (2018) also compare the performance of wholly government-owned VCs to privately-owned VCs in China.  They find that companies backed by government-owned VCs have a lower probability of achieving an IPO than those backed by privately-owned VCs.

Three other studies shed light on the impact of VC-investing in China. Guo and Jiang (2013) find that VC-backed firms outperform non-VC-backed firms in terms of profitability, labor productivity, sales growth, and R&D investment. Their evidence suggests that this outperformance is driven by both superior project selection and the ex-post monitoring efforts of VCs. Humphery-Jenner and Suchard (2013) examine the role of foreign VCs in venture success for Chinese portfolio firms. They find that foreign VCs do not increase the overall likelihood of venture success. They also find that foreign VCs seem to prefer exiting via a M&A or a secondary buyout rather than an IPO, perhaps because foreign VCs lack the local knowledge and political connections useful in securing the latter. Zhang et al. (2017) provide further evidence that well-connected VCs can be useful in navigating the IPO process. Specifically, the authors show that late (pre-IPO) stage VC investors earn higher returns than those that enter in the early stages. Furthermore, this result is more prominent for VCs with strong political connections. This finding suggests that the uncertainty associated with China's IPO approval process can impose real costs on startup businesses, and that these businesses may seek to mitigate such costs by engaging well-connected VC investors.

An important recent development has been the emergence of two dominant corporate investors. In the past ten years (2008-2017), Alibaba and Tencent have together invested more than US$ 46 billion in Chinese startups. Table 3 reports the value of the investments these two firms have made each year (extracted from PEdata). Panel A reports only investments made through their VCPE funds; Panel B also includes strategic investments made through other corporate entities within each group.[16] This table shows that these two firms have each invested around US$23 billion in Chinese startup firms. Most of the investments (US$18.6 billion for Tencent and US$15.7 billion for Alibaba) were made through their VCPE funds, but a sizeable minority was made as strategic investments through other corporate entities. The total VCPE investment of these two firms (US$34.33 billion, in Panel A) represents 4.90% of all the Chinese VCPE investments over this period. Clearly these two leading names are leveraging their domain

---

[16] The Tencent group invests through four corporate entities (Tencent, Tencent Linzhi, Tencent Cloud, and Tencent Music) and three VCPE funds (Tencent Capital, Tencent Startup Space, and Tencent Venture). The most important vehicle is Tencent Capital, which accounts for more than 90% of the group's investments in Chinese startups. The Alibaba group invests through 6 firms and 2 VCPE funds. The most important vehicle is Ali Capital, which accounts for more than 90% of the investments made by the Ali group.

knowledge and access to financial market, and playing important roles in venture financing for Chinese entrepreneurs.

Although China is transitioning towards a market-based economy, many aspects of the governing apparatus still bear the hallmarks of "central planning." The most prominent example is the central government's "Five-Year Plan". Of particular interest is whether the government actions outlined in these plans lead to better resource allocation in the domain of entrepreneurial finance. Zhou et al. (2017) examine how policies in China's 12th Five-Year Plan, published in 2010, affected the performance of VC funds. This particular Five-Year Plan identified seven strategic emerging industries (SEIs), and outlined a series of policy measures designed to facilitate the development of these industries.[17] Interestingly, Zhou et al. (2017) find that these policies significantly *reduced* the exit performance of VCs that focused on the targeted industries. Furthermore, the erosion in performance is concentrated in privately-owned VCs and is not evident in government-owned VCs. Evidently, government involvement in the form of this edict distorted market incentives, leading to over-allocation of VC resources to the targeted industries. This finding supports Ahlstrom, Bruton and Yeh's (2007) view that the government involvement can often have unintended negative consequences for VC firms and entrepreneurs.

Taken together, the studies above suggest that in China, as in other developed countries, VC participation is associated with improved entrepreneurial performance. Some amount of government ownership, in particular, can be helpful when exiting through an IPO. Furthermore, the evidence suggests that government involvement through central planning edicts, while well-intended, may lead to worse resource allocation decisions by the participants in the domain of entrepreneurial finance.

---

[17] The seven SEIs listed in this plan are: energy efficient and environmental technologies, next generation information technology (IT), biotechnology, high-end equipment manufacturing, new energy, new materials, and new-energy vehicles (NEVs).

*3.3 Late-stage Financing*

The central issue in late-stage venture financing is the exit strategy, and it is in this area that we find the biggest opportunity for public policy improvement in China. The exit options that entrepreneurs and the venture investors have available to them clearly impacts their willingness to engage in such ventures in the first place. Because they impose risks and costs on financiers, regulatory delays and other exit impediments can increase the cost-of-capital to entrepreneurs. In this subsection we provide some preliminary evidence on how venture investors currently exit their positions in Chinese startups. The findings presented here lead naturally to policy issues surrounding China's IPO regulations, which we discuss in more detail in Section 4.

Table 4 presents the means of exit for VCPE investors in Chinese startup firms. The data is extracted from PEdata. Each investor-investee transaction is a separate event in this database. An exit event is defined as the unwinding of a transaction previously entered in the database as an investment event. Panel A reports a breakdown of all exit events by category: IPO denotes initial public offerings, Repurchase denotes shares reacquired by the investee, M&A is merger and acquisition, and Equity Transfer refers sales from one investee to another. Panel B further breaks down the IPO exits by stock exchange. In this panel, SHSE is the Shanghai Stock Exchange and SZSE is the Shenzhen Stock Exchange excluding the ChiNext Board. Although the ChiNext Board is officially part of the SZSE, it is separately reported here because listing requirements are lower and it is an important exit market for venture investors. US IPOs include listings in the NYSE, AMEX, and the NASDAQ. Hong Kong IPOs only include listings on the main board as the Growth Enterprise Market (GEM) is extremely illiquid.

Panel A of Table 4 shows that the most common means to exit is via an IPO (4,415 events, representing 43.1% of all exited deals), followed by a M&A transaction (28.9% of all exits) and an equity transfer (20.7% of all exits). Panel B of Table 4 further breaks down IPO exits by exchange. The domestic exchanges (SHSE, SZSE, and ChiNext) collectively account for 3,044 IPO exit events, representing 76% of total IPO exits. Listing on the U.S. and Hong Kong exchanges together accounted for 956 exit events, representing 23.8% of total IPO exits.[18] These results show that VCPE funds investing in Chinese firms rely heavily on IPOs, M&A, and equity transfers to close their investment cycle. Domestic IPOs, in particular, account for over 30% of the exit events that took place over the past 12 years.

Table 5 presents the number and dollar value of Chinese firms that undertook an initial public offering (IPO) or a reverse merger (RM) transaction in the years 2007 to 2017. All monetary amounts are in billions of U.S. dollars. We include all firms headquartered in China that went public through an IPO in China Mainland, Hong Kong, or the U.S. For each IPO, we report the total funds raised, as well as the implied market valuation for the firm as a whole based on the IPO issue price. We also include firms that accessed the China Mainland market using a RM transaction as per Lee, Qu, and Shen (2018a, 2018b; hereafter LQSa and LQSb).[19] For each RM, we report the market value at the announcement of the RM transaction, as well as the current market value of the firm based on closing prices as of the end of March, 2018. The IPOs in mainland China include the IPOs in Shanghai and Shenzhen Stock Exchange. The data source is CSMAR. The Chinese firms IPOs in Hong Kong include the IPOs in Hong Kong Main Board. The data source is WIND and CSMAR. The Chinese firms IPOs in U.S. include the IPOs in AMEX, NYSE and NASDAQ. The data source is Compustat and SDC Platinum.

One striking fact from Table 5 is the relatively small sum raised through Mainland IPOs. This is because the larger Chinese firms are choosing to list overseas or to circumvent the Mainland IPO process by listing through a RM. The lethargic long-term performance of the A-shares market further exacerbates the issue (Allen et.al. (2014)). Table 5 shows that while most Chinese firms IPO in the mainland (2,087 Mainland IPOs, versus 582 listings for Hong Kong and U.S. combined), the mainland IPOs are much smaller. In fact, over the sample period mainland IPOs actually raised less money than the oversea IPOs (US$ 50.32 billion, versus US$ 66.64 billion for Hong Kong and U.S. combined). Furthermore, consistent with Lee, Qu, and Shen (2018b), we find that domestic RM are also much larger than domestic IPO firms. RMs are not financing events, so we cannot compare the amounts raised. However, Table 5 shows that the RMs begin their lives at an average market capitalization of US$ 2 billion each, while the average IPO is valued at US$ 0.194 billion each (406.44/2087) based on issue prices. Even allowing for a 100% IPO initial day underpricing, the typical mainland IPO is an order of magnitude smaller than the typical RM. The initial market valuation of the 133 RMs is US$266.6 billion, which is 65.6% of the market valuation of all the 2,087 mainland IPOs combined. These results are quite striking given the fact that oversea listings and RMs are quite costly to the issuing firms.[20] Evidently many large Chinese firms are willing to incur real costs to avoid the arduous mainland IPO process.

---

[18] According to the PEdata, the listing in exchanges is counted as an exit event except the listing in NEEQ in which VCPE investors need to sell their shares. As discussed later, the listing in NEEQ usually is not a financing event.

[19] Following Lee, Qu, and Shen (2018a, 2018b), our RM sample includes all "clean RM transactions," wherein control of the combined entity is transferred to the owners of the unlisted firm, and the entire exchange takes place in a single transaction.

Table 6 provides additional insights on the type of firms that choose to IPO in a Mainland exchange or overseas, as well as those that elect to access a Mainland exchange through a RM transaction. Specifically, we report the total assets (Assets), return-on-assets (ROA), market-to-book ratio (MB), book leverage (Lev) and cash holdings divided by total assets (Cash). All financial numbers are as of the last pre-listing report. See accompanying text in the table for details on variable construction. Panel A reports the mean for each variable; Panel B reports the median. Bold fonts denote statistical significance at 1% in a difference-in-mean (Panel A) or a difference-in-median (Panel B) test between the Mainland IPO firms and the firms identified in each column heading.

This table highlights a number of differences between firms that go public through a mainland IPO and those that listed through other means. Many of these differences reflect the expressed preference of the CSRC IPO review committee for stable, profitable firms that have solid balance sheets with higher tangible assets. Note in particular that Chinese IPOs in the U.S. are significantly less profitable (lower ROA) but have much higher market-to-book ratios (MB).[21]

---

[20] Lee, Qu, and Shen (2018a) report that the typical shell value paid in a mainland RM transaction is around US$ 400 million.

[21] In computing the market-to book (MB) ratios, we use the implied market value of the firm as a whole based on the IPO issuance price. The CSRC has an unwritten requirement that the maximum PE ratio for a mainland listing be

They have less total assets (Assets) but much more of their asset is in the form of cash holdings (CASH). Indeed, compared to their Mainland counterparts, the Chinese firms that IPO in the U.S. are much more likely to be fast growing technology firms. The firms that IPO in Hong Kong seem closer in profile than those that IPO on the mainland, but we still see significantly higher leverage and lower cash holdings than the mainland IPOs. Consistent with LQSb, we find the mainland RM firms are larger, less profitable, more levered, and hold less cash than the mainland IPOs.

Finally, Table 7 presents an overview of the National Equities Exchange and Quotations (NEEQ) Market. The NEEQ was established in 2013 as an attempt to provide more visibility to early startup enterprises in China. This market operates in a manner similar to the pink sheet over-the-counter system in the United States. Listing on the NEEQ is not a financing event and liquidity is generally quite thin. In this table, total market capitalization and total trading volume are in billions of U.S. dollar. In Panel A we provide summary statistics on market liquidity. Total trading volume is further subdivided into trades that involve a market maker, and trades that were consummated through bilateral negotiation between the parties. In Panel B, for comparison, we also present market liquidity statistics for the A-Share markets (Shanghai and Shenzhen combined). In Panel C we report a distribution of the number of investors for the NEEQ firms, as of the end of 2017.

Table 7 highlights the thinness of the NEEQ market. As of the end of 2017, 11,630 firms are listed on the NEEQ, with a combined market value of $US 760.07 billion (or 8.74% of the market capitalization of the main A-Share markets). Note that the NEEQ market value is likely overstated, as price discovery is minimal and share prices are in some cases self-reported. A better measure of the economic scale of this market is its total trading volume. In 2017, total NEEQ volume is US$ 34.95 billion, which is a tiny fraction (0.21%) of the volume on the main A-Share markets. Furthermore, most of this volume was generated by trades involving bilateral

---

limited to 23 (see Section 4 for details) when establishing the listing price. As a result, the MB ratios for mainland IPOs are artificially depressed. However, even if we assume the issue price is allowed to double (i.e., if the maximum PE ratio is raised to 46), the MB ratios in mainland IPOs are still much lower than those in U.S. IPOs. This difference most likely reflects the higher growth potential associated with U.S. IPOs.

negotiation between the buyer and the seller (essentially "trades by appointment"). The actual volume that involved market makers is only US\$ 12.07 billion. Panel C further shows that 83.6% of all the firms on the NEEQ have 50 or fewer investors. Only 3.45% of these firms have more than 200 investors. In untabulated results, we also find that the accounting qualities of NEEQ firms are significantly lower than those in the main A-share markets. In short, while the NEEQ currently provides some visibility to a large number of startup firms, it is unclear how useful this visibility is in resolving the financing needs of these startups.

The analyses in Section 3 highlighted the surprisingly small role that mainland IPOs have played in the exit strategies of the venture financiers investing in Chinese startups. More VCPE investments have been closed through M&A and Equity Transfers than through IPOs (Table 4). For firms that do elect to go through an IPO process, more money has been raised overseas (in Hong Kong and the U.S.) than in the mainland. Furthermore, a disproportionate number of high-growth technology firms are listing overseas rather than in the mainland. All these results point to potential problems with the mainland IPO approval process, a subject that we turn to in the next section.

## 4. China's Initial Public Offering (IPO) Policies

The literature on law and economics has nominated two main competing theories on the role of laws and institutions in the functioning of markets. *Public interest theory* (Pigou 1938) holds that unregulated markets are prone to failures, and that regulations help protect investors from market failures, such as unscrupulous operators and various negative externalities. Conversely, *public choice theory* (Tullock 1967; Stigler 1971; Peltzman 1976) sees the government as less benign, and regulation as socially inefficient. In one particular form of public choice theory, the regulatory agency is "captured" (Stigler 1971) by industry and operates primarily for its benefit. In an alternative form, the regulatory agency operates a "tollbooth" primarily for the benefit of bureaucrats and politicians (McChesney 1987; De Soto 1989; Shleifer and Vishny 1998).

The IPO regulations in China present an interesting setting in which to examine the predictions of these competing theories. One the one hand, the rationale for China's heavy IPO regulation can be framed in terms of Pigou's *public interest theory*. Given her weak corporate governance structure and limited investor recourse through the court system (e.g., see Allen et al., 2005; Jiang, Lee, and Yue 2010), China may need stringent entry rules to protect investors against unscrupulous operators seeking access to public markets (e.g. Pistor and Xu 2005; Du and Xu 2009). On the other hand, these stringent rules may be better understood in the context *public choice theory*, wherein they exist and are maintained mainly for political/industrial purposes. Under this theory, the regulations actually provide minimal benefit to the investing public and are kept in place primarily because they can bestow economic rent upon politicians, bureaucrats, and industry incumbents (e.g., Shleifer and Vishny 1998).

The economic calculus of regulation calls for a policymaker to weigh the benefits of a rule against its costs. While it is difficult to quantify the benefits of China's IPO policies, in terms of investor protection, the same cannot be said of its costs. In fact, a large literature now discusses the economic problems and consequences directly attributable to the country's current IPO regulations. In this section, we survey this evidence and offer our own commentary. We organize this review around nine stylized facts that have emerged from this literature.

*1. Long wait times and substantial outcome uncertainty for candidate firms*

After 2013, the CSRC posted on their website information for each candidate firm that has been reviewed. Lee, Qu, and Shen (2018a; LQSa) manually collected this data and studied success rates and waiting times. Their results show that between 2014 and 2017, a total of 333 candidate firms received a final decision. Among them, 113 (34%) were rejected and 220 (66%) were approved for listing. The average waiting time (from the initial submission to the listing date) for the 222 firms that received approve is 24.5 months. LQSa also notes that historical wait times prior to 2014 are likely higher, due to several extended suspension periods.

In sum, first that submit their IPO applications to the CSRC can expect to wait between one to three years before their files are reviewed, at which time they have a one in three chance of being approved. In our discussions with Chinese VCPE firms, we learned that during the wait period it is not advisable for the candidate firm to change their profile substantially (for example develop new product lines, create new markets), as these changes can jeopardize their candidacy. Essentially, candidate firms need to press the pause button on many aspects of their operations after filing an IPO application.

*2. A bias against high-growth technology firms*

IPO qualifications on Mainland exchanges are tilted heavily in favor of stable firms with steady earnings and conservative balance sheets, and against firms that invest heavily in R&D or have substantial intangible assets. This bias is evident in the pre-specified requirements that must be met by IPO candidates even before they register in the CSRC queue.

For example, minimum requirements for IPO registration on mainland exchanges, as stated in the 2006 directive issued by the CSRC, include the following bright-line thresholds:

1. The firm must have been in operation for more than 3 years.
2. There must be no major change in the main business operation, directors, top executives, or controlling shareholders in the past 3 years.
3. The net income in each of the past 3 years must be positive, and the cumulative net income in past 3 years must be more than 30 million RMB.
4. The cumulative operating cash flows in the past 3 years must be more than 50 million RMB; or the cumulative gross revenues in the past 3 years must be more than 300 million RMB.
5. The total equity before IPO must be more than 30 million RMB.
6. The ratio of intangible assets to net assets should be less than 20% in the most recent years.
7. The firm does not have any unrecovered losses in the most recent year.

These requirements would have disqualified many of the most innovative and impactful technology companies that publicly listed in the U.S. in recent years, including LinkedIn, Twitter, Tesla, Dropbox Inc. and Spotify Technology.  In fact, Eisen (2018) report that more than three quarters of the 108 companies that completed IPOs in the United States in 2017 reported losses in the 12-months leading up to their public listing date.  None of these companies would have been allowed to list under current Chinese IPO regulations.

The recent establishment of the National Equities Exchange and Quotations (NEEQ) is a partial response to this problem.  The NEEQ (often referred to as "The New Third Board") officially opened on January 16, 2013.  Its main function is to facilitate the trading of shares in public limited companies that are not listed on either the Shenzhen or Shanghai stock exchanges.  As discussed in Section 3, over 11 thousand companies are now registered with the NEEQ. However, as our results show, the NEEQ is thinly-traded and most of the trading that takes place arises from bilateral negotiations between buyer and seller rather than from market making activities.  More importantly, although registration on the NEEQ may help increase visibility, it is not a financing event.  Given the illiquidity of this market, the NEEQ cannot be viewed as an effective exit option for entrepreneurs and early investors.

*3. Substantial underpricing of IPOs*

Mainland IPOs are typically accompanied by outsized initial-day returns.  For example, using a sample between 1992 and 2004, Tian (2011a) shows that the average (median) first day return is 247% (122%).  Between 1999 and 2007, Chen et al. (2015b) find that the average (median) first day abnormal return is 127% (105%). The differences across studies reflect different sample periods and as well as some changes in the regulations governing IPO offer prices (Chen et al., 2016a). Using a recent sample from 2007 to 2015, LQS (2018b) still find an average (median) initial day abnormal return of 117% (53%).

These numbers are much larger than the initial-day returns associated with IPOs in most other countries.  For example, Loughran and Ritter (2004) show that initial day return in U.S. is 7% in

1980s, 15% during 1990-1998, 65% during the bubble years of 1999-2000, and 12% during 2001-2003. Using an international sample of 37 countries from 1998 to 2008, Boulton et al. (2011) show that, excluding China, the average IPO underpricing ranges from 1.43% in Argentina to 58.33% in Greece.

The IPO underpricing effect in China is attributable largely to an unwritten rule that calls for issuing firms to cap the maximum price-to-earnings (P/E) ratio when establishing an issue price. Although the rule is not codified, it is widely acknowledged by all parties. Our discussions with investment banks, issuers, and venture capitalists all confirm that in recent years, it is well understood that the proposed issue price should not exceed 23 times the firm's "normal" earnings. Firms that price their shares higher than this ratio risk jeopardizing their chances of an approval.

One clear implication of these results is that electing to IPO on a mainland exchange can be quite costly to the issuer. Although investors with pre-listing allocations earn large profits, the issuing firm often receives less than half of the total wealth generated through the listing process. In our view, assuming full disclosure of relevant facts, firm value assignment and adjudication is better done by market participants rather than regulators.

*4. An exodus of high-quality technology firms overseas*

Section 3 results show that over the past 11 years, total financing received by Chinese firms that IPO in Hong Kong and China exceeded the financing received by firms that IPO in the mainland. Strikingly, these results also show that at the time of their IPO, the oversea Chinese firms were less profitable, more levered, and have much higher market-to-book ratios. Many of these are high-quality technology firms (most notably Alibaba and Tencent, and more recently Sogou Inc.).

It should come as no surprise that higher quality technology firms want to be recognized as such, and to be weighted on a more equitable scale. To the extent that these firms believe the value assigned to them by investors in a foreign exchange better reflect their true worth, they would naturally migrate oversea.[22]

*5. Costly reverse merger (RM) transactions*

Another way to circumvent the mainland IPO process is through a reverse merger (RM). Lee, Qu, and Shen (2018a; LQSa) compiled a comprehensive sample of 134 "clean RM" transactions that took place on mainland exchanges, wherein control of the combined entity is transferred to the owners of the unlisted firm, and the entire exchange takes place in a single transaction. Their results show that during the 2007-2015 period, unlisted Chinese firms paid an average of 3 to 4 Billion RMB (more than $400 Million USD) for each listed shell, an amount exceeding 2/3 of the median market capitalization of listed firms at the time of the RM.

Because these shell values are established in arms-length transactions between listed and unlisted entities, LQSa argue that they provide a "shadow price" for the cost of accessing China equity markets. One immediate implication of these large shell values is that the "gate fee" associated with entry into Chinese equity markets is extraordinarily high. These large shell values show that, in China, a RM is not a low-cost channel through which to become publicly-listed. The high price paid for shell companies also raises the specter that Chinese IPO regulations may be too stringent and are socially inefficient. Rather than serving as an effecting screening mechanism for weeding out lower-quality firms, these policies may in fact be preventing healthy firms from accessing public equity markets.

Picking up on this theme, LQSb study the choice to go public through RMs versus IPOs in mainland markets. Their results show that pre-listing RM firms are larger, more profitable, but less politically-connected than IPO firms. Chinese RM firms also have superior post-listing performance, both in terms of operations and in terms of stock returns, compared to IPOs matched on industry and size. These results are in sharp contrast to the evidence on RMs from developed countries, where RMs are smaller, riskier, and lower-quality "backdoor listings." Again, the evidence suggests China's stringent IPO policies are blocking high-quality firms from accessing domestic equity markets.

---

[22] One reason Alibaba gave for its decision to list in the U.S. is that investors and analysts "better understand" their business model and value proposition.

*6. Virtually no delisting or retirement of failed companies*

In the normal course of events, listed firms that perform poorly should become involuntarily delisted. However, given China's high shell value, one might expect failing firms to find other uses for their coveted listing status. LQSa show that this is indeed the case. Specifically, the authors find that the average mortality rate for Chinese listed firm (defined as an involuntary delisting for performance reasons) is only 0.6%, or 1/40[th] the mortality rate for U.S. listed firms over the same period. The few firms that were delisted typically involved fraud or criminal behavior. Essentially, publicly-listed Chinese firms almost never delist for performance reasons (Allen et.al. (2014) also discusses this phenomenon).

Instead of delisting or becoming a shell, poorly-performing firms can exploit their listing status by taking over another unlisted business. LQSa show that rather than surrendering their listing status through a RM transaction, many floundering businesses "reinvent" themselves through a major asset restructuring (MAR).[23] In fact, they show one in five (19%) of the firms in their highest Estimated Shell Value (ESV) decile portfolio – i.e., those firms most likely to become shell candidates – actually undergo a MAR in the next 12-months. Evidently many prime shell targets are electing to retain control rather than surrendering their listing status. To the extent these floundering firms do not have the best managerial talent, these findings point to a further economic cost of the current IPO regulations – namely, the ability of poor managers to prolong their tenure by leveraging their listing status through MAR transactions.

---

[23] The CSRC defines a MAR as an event where more than 50% of a firm's operating assets has changed, and requires firms engaged in a MAR to report such transactions. Asset swaps that do not result in a change in control will not give rise to a RM, but if the scale of the asset swap is large enough, it will result in a MAR filing.

*7. Large cross-sectional price distortions among listed stocks*

Given an average shell value of US$400 million, at any given time around 30% of China's publicly trades stocks will be trading below their shell value. This should lead to a number of predictable price distortions among listed stocks. LQSa hypothesize and document several such effects:

a. *Expected shell value (ESV) is a priced factor.* The authors define each firm's expected shell value (ESV) as the product of the average prevailing shell value and the predicted probability of shell (ESP), divided by each firm's market equity. They then examine the returns to a hedged-ESV portfolio that is long (short) firms with the highest (lowest) ESV. Their results show that high ESV firms earn higher future returns, even after controlling for other firm characteristics known to be associated with expected returns in China. Specifically, they find that a "long-short strategy based on extreme ESV deciles generates a raw return of 29% per annum and an abnormal return of 5.4% per annum after controlling for all five Fama-French factors. These findings establish the 'shell effect,' or the shell premium, as one of the most prominent predictive variables for equity returns in the Chinse stock market." (LQSa, p.4).

b. *ESV returns are sensitive to IPO-related regulatory shocks.* If RMs are driven by IPO rationing, returns to the ESV factor should be responsive to regulatory shocks associated with policy changes affecting IPO and RM activities. In particular, LQSa posit that regulatory changes tightening (relaxing) the IPO quota would be associated with positive (negative) returns to the hedged-ESV portfolio. Conversely, regulatory changes that tighten (relax) access to RMs would be associated with negative (positive) returns to the hedged-ESV portfolio. Using a set of six event studies, they find that this is indeed the case. Specifically, the shell premium (i.e., returns to EVS factor portfolio) increases sharply around the announcement of tightened IPO regulations. The opposite result is observed around events that tightened RM regulations. These results support the view that the ESV factor captures cross-sectional differences in firms' sensitivity to IPO-related regulatory risk.

c. *High-ESV firms are insensitive to corporate earnings.* If high-ESV firms trade primarily on their expected shell value, their stock price should be less sensitive to corporate earnings. LQSa find this is indeed the case: for high-EVS firms, they find a negative relation between reported earnings and the market-to-book ratio, which is in sharp contrast to the positive relation that exists between earnings and pricing multiples for low-ESV firms. This result again suggests that China's IPO policies have a distortive effect on the stock returns of its publicly listed firms.

Taken together, these findings suggest that regulatory risk, as captured by fluctuating shell values, is an important economic driver of cross-sectional returns (particularly the large returns earned by small firms) in China.

*8. Listing delays leading directly to a reduction in innovation activities*

The IPO approval committee at CSRC controls the aggregate approval rate for IPO applications based on market conditions, and out of concern that too many IPOs may lead to reduced liquidity and market valuation (Braun and Larrain, 2008); Tian, 2011a). One extreme form of regulation that delays listing is an IPO suspension, which puts a pause to all IPO activities beyond the application submission step. Between 1994 and 2016, there have been nine major IPO suspensions lasting 3 to 14 months. The start and end of the suspensions were often determined at ad hoc meetings and were not announced beforehand, therefore were not fully anticipated by market participants. Such delays are especially costly to firms already approved to IPO because of foregone strategic opportunities (e.g. to make acquisitions or large investments using public funds) and disruptions to long-term corporate plans to invest and innovate.

Cong and Howell (2018) use a novel identification strategy to examine the effect of a listing delay due to China's IPO policy on firm's innovation-related activities. Specifically, they assemble a comprehensive data set from multiple sources including IPO prospectuses, Google Patents, CSRC, and State Intellectual Property Office, covering 1,567 firms that were approved and ultimately went public on the Shanghai and Shenzhen exchanges between 2004 and 2015.

Focusing on two recent suspensions affecting firms approved with in the year before each suspension, they exploit the fact that the assignment to the control (firms listed right before the suspensions) and treatment groups (firms listed after the suspensions ended) is stochastic and not driven by firm-specific factors.[24] The authors argue that suspension-affected firms faced a longer time between approval and listing as well as greater operational uncertainty, and then examine the lasting impact of temporal delays to accessing the public markets.

Their results show that suspension-affected firms, which experience on average 16 months of delay, have 28 percent fewer Chinese patent applications than other firms, which spend on average three months from listing approval to actual listing. Suspension-induced delay also reduces patent quality; in the year after IPO approval, there are large significant declines in granted Chinese patents, citations to Chinese patents, and granted global (non-Chinese) patents. The effect persists for multiple years. For example, four years after the approval year, the treated group has 16 percent fewer patent applications.

The authors also find a positive treatment effect on leverage, a negative effect on tangible investment and return on sales, and no effect on total sales or earnings, but none of these effects lasts beyond the first year of the IPO. Taken together, their findings suggest that innovation is a cumulative process and that delays in the process can have lasting effects on outputs. They also establish a direct link between frictions in the financing process and output-based measures of firm innovation. Specifically, these findings provide an estimate for the economic loss, in terms of innovation-related outputs, associated with the detrimental listing delays. In China where public market provides important risk capital for innovation, the importance of timely access to a well-functioning market cannot be overemphasized.

*9. Potentially inflated prices and excessive speculative trading*

---

[24] Even though application approval could be driven by specific firms' connections with local politicians (e.g., Piotroski and Zhang 2014), the CSRC and local governments do not intervene in the listing process due to firm-specific reasons (except for illegal activities) once the application is approved.

China's restrictive IPO regulations are also associated with another problem: the possibility that stock prices in domestic markets can drift further away from fundamental values. These entry regulations form a costly barrier, insulating the domestic market from arbitrage forces at work in other markets. As a result, institutional ownership in mainland markets is low, and foreign ownership is even lower, leading to much noisier stock prices.[25] This effect can be seen in two related empirical phenomena: (a) a wider corridor for mispricing relative to fundamentals, and (b) greater levels of speculative trading.

Evidence of the mispricing precipitated by the artificial entry barrier can be seen in the pricing of cross-listed stocks. A number of Chinese firms are cross-listed in both mainland and Hong Kong. For many years the price of A-shares routinely diverged from (and is typically higher than) the price of H-shares for the same company traded in Hong Kong, even though these shares are equivalent claims on the same underlying cash flows. A useful measure to gauge the price disparity is the Hang Seng Stock Connect China AH Premium Index ("HSAHP"), which tracks the average price differential of A shares over H shares for the most liquid cross-listed companies. When HSAHP is higher (lower) than 100, A-shares are trading at a premium (discount) relative to H shares.

Over the past 11 years, the annual average HSAHP ranged from a low of 97.98 in 2014 (the only year below 100), to a high of 148.37 in 2007. In earlier years, the premium tended to be larger. For example, Wang and Jiang (2004) study a sample of 16 A-H duel listed stocks from 1996 and 2001 and find that the average daily price premium of A- relative to H-shares is about 32%. In recent years, the pricing differential appears to have decreased. However, even as of this time of writing, a sizeable A-share premium remains. The average daily HSAHP for 2018, as of August 17, is 123.40, suggest an average premium for A-shares of approximately 23%. The persistence of this pricing difference speaks both to the limits to arbitrage at play, and the general tendency for domestic Chinese investors to prefer A-shares.[26]

---

[25] For example, Jiang, Lee, and Yue (2010, Table 8) report that for stocks listed on mainland exchanges, the mean (median) ownership by institutional investors in 2004 was only 8.64 (1.26) percent of the total tradable shares. They argue that this low level of institutional ownership is one reason why stock prices did not fully incorporate the value implications of self-dealing (or "tunneling") by firms' block shareholders. Institutional ownership has increased in recent years, but it is still much lower than in other developed countries.

Another by-product of market fragmentation traceable to the insular nature of mainland markets is higher turnovers. Pan et al. (2016) report the annual turnover rate of twenty largest equity markets worldwide at the end of 2012. In the 1990s, the annual turnover rate in China is nearly 500%, while the annual turnover rate for other countries rarely exceeds 100%. After 2000, the annual turnover rate in China gradually decreased to 200%, but is still much higher than those in other countries. Pan et al. (2016) attribute this high turnover rate to speculative trading and show that a measure of abnormal turnover rate negatively predicts future market-wide returns in China. Similarly, Hu et al. (2018) report the monthly turnover in mainland markets averaged around 20% during 1990 to 2015. They find that the monthly turnover exhibits wide fluctuations over time without any obvious time trends. For example, China's monthly turnover rate exceeded 120% in 1994, 1997 and 2007, and dropped below 10% in 2002, 2012 and 2013. These wide swings seem to reflect ebbs and flows in the level of retail investor interest in mainland equities. To the extent that the institutional appetite for equities is different, policies that invite greater institutional participation (particularly by foreign investors) should lead to more stable domestic markets.

## 5. Summary

### 5.1 Reflections on Public Policy Implications

---

[26] Recent innovations, such as the initiation of the Mainland-Hong Kong Stock Connect program in 2016, could reduce arbitrage costs over time. In brief, this program allows mainland investors to buy/sell Hong Kong listed shares, and Hong Kong investors to buy/sell mainland shares. However, a number of factors make this trade far from riskless. First, currency is an issue, as A-shares are priced in yuan and H shares in Hong Kong dollars. Second, investor preference in the mainland still favors A-shares. Even when the same company's H-shares are available at a cheaper currency-adjusted price through Stock Connect, not enough mainland investors are selling their A-shares in favor of the Hong Kong offering. Third, only larger and more liquid mainland stocks are available through Stock Connect, and it is extremely difficult to short sell mainland shares. Finally, although both types of shares are legal claims to the same underlying cash flows, no direct arbitrage exists. This is because A- and H-shares of the same company cannot be bought and sold in each other's respective markets. All these factors conspire to generate the 20+% premium we observe today. Other studies that examine the A/H pricing puzzle include Chan et al. (2009), Zhao and Liu (2005), and Zhang and Zhang (2018).

In this study, we have reported on the current state-of-affairs in the funding of entrepreneurship and innovations in China and provided a broad survey of academic findings on the subject. Rather than focusing narrowly on a specific subject in this domain, our goal is to curate a broad set of related results and present them for evaluation from a public policy perspective.

Our analyses reveal an exponential rate of growth in total VCPE funding for "mid-stage" enterprises operating in China, particularly over the most recent three years. In 2017, total VCPE funding for Chinese firms exceeded US$400 billion, and is second only to the United States. We show that most of this funding is coming from Chinese rather than foreign sources, with both government entities and Chinese private corporations playing substantial roles. This surge in mid-stage funding has occurred in concert with, and endogenous to, the rise of many highly innovative private technology firms, each with a valuation above US$ 1 billion (the 164 Chinese "unicorns" mentioned at the outset). The level of innovative activities in China, as measured by these metrics, has never been higher.

At the same time, our analyses suggest opportunities for improvement in the "early-stage" of the entrepreneurial life cycle. Compared to their counterparts in the United States, early-stage entrepreneurs in China appear to have less access to microfinancing, crowdfunding, accelerator, and incubator programs (with China's "Innofund" being a notable exception). Beyond financial support, early-stage entrepreneurs often need advice, mentorship, connections, and other educational components not currently available from programs such as the Innofund. For example, in Canada, a network of government organizations such as the Ontario Network of Entrepreneurs (ONE), working together with regional not-for-profit organizations such as Haltech, MaRS, Communitech, and Invest Ottawa, have drawn international attention. These entities are staffed with advisors and mentors who help startups commercialize their ideas. According to a fall 2017 report on the ONE, this organization alone helped more than 5,600 Ontario entrepreneurs to open new businesses (Khorsravi 2018b). This would appear to be a potentially interesting direction for Chinese policy makers to consider.

Our analyses also point to opportunities for improvement in the "late-stage" of the entrepreneurial life cycle. Specifically, we find support for the view that the current regulations governing the initial public offering (IPO) are antiquated and in dire need of reform. These rules require candidate firms to endure long wait times, with a significant risk of delays and rejections. The current rules also bias against high-growth technology firms, which typically have lower profits, less developed businesses, and more intangible assets. Furthermore, a growing body of academic evidence now links the outcome of the IPO review to a candidate firm's political-connectedness.

Given the problems associated with domestic IPOs, it is not surprising that many Chinese entrepreneurs have sought alternative exit strategies. Our results show that over the past 11 years, more money has been raised by Chinese firms in oversea IPOs (in the U.S. and Hong Kong) than in mainland IPOs. Alternatively, to avoid the IPO gauntlet, many private firms have spent an average of US$400 million each to purchase a "shell" company in a RM transaction. These large shell values have in turn led to price distortions in China's equity markets, include extraordinarily large returns to the smallest firms. The large shell values also help to explain why failed companies almost never delist from public equity markets. In fact, these failed businesses continue to propagate by levering their listing status to acquire new lines of business, thus maintaining control despite a demonstrated track record of failure. At the same time, the private firms that are absorbed by these failed entities are circumventing the IPO process. As a result, a significant number of unlisted businesses are gaining public market access without having undergone the mandatory scrutiny and review.

In our view, China's current IPO regulations represent the single most serious impediment to two important near-term goals expressly espoused by the Chinese government: (a) to bring more high-technology firms back to mainland stock markets, and (b) to be included at a meaningful weight in international stock indices, particularly the MSCI Emerging Market Index. We discuss each of these goals in turn.

In March 2018, the State Council passed new regulations allowing for fast track approval of Chinese Depository Receipts (CDRs) by technology firms that have listed overseas. The CSRC is also actively lobbying large China private technology firms to list domestically rather than abroad (e.g., Yu and Zhang (2018)). No doubt some firms will respond to the "expedited IPO processing" offers now being extended by Chinese regulators.  But others may see these policies as addressing the symptoms rather than the root cause.  Theory suggests higher quality firms will want to be recognized as such.  To the extent that these firms believe a foreign listing better signals their worth, or foreign markets offer greater depth and liquidity, they may still choose to list oversea.   In our minds, a fundamental overhaul of the IPO review system provides a clearer and more credible signal to these private technology firms that Chinese equity markets will provide a long-term solution to their financing needs.

With respect to the second goal, China A-share is currently the second largest market in the world; yet it is also the most under-owned (by foreign investors).  A key reason for the under-weight is A-share's exclusion by closely watched and trusted index providers like MSCI Inc.  In 2013, MSCI put A-shares on a review list but declined to include them in any indexes, citing issues including capital mobility restrictions and uncertainties around taxes.  It continued to reject A-shares in 2015 and 2016.  Finally, in June 2017, MSCI announced a 5 percent partial inclusion to be implemented in 2018.  In doing so, MSCI expressly acknowledged China's efforts to reform its capital markets.  The partial inclusion means A-share stocks will form roughly 0.73 percent of the MSCI Emerging Market (EM) index and 0.1 percent of the MSCI All Country (AC) World index by the end of 2018.  Analysts estimate this partial inclusion translates into $20 billion in foreign investment in Chinese markets.  At full inclusion China would be around 20 percent of the MSCI EM index.  This would translate into at least an additional $300 billion in foreign investments.  However, MSCI is clear that further inclusion, if any, would depend on the pace of market reform in China.

Considering the problems identified above, we recommend a move toward a registration-and-disclosure system for Chinese IPOs, like those employed by most other countries.  In such systems, investors monitor firm quality and market forces adjudicate firm value.  Firms that receive enough support from the investment community will attain IPO status.  The role of regulators is to ensure adherence to established ordinances, which are largely disclosure-centric.

We acknowledge some concerns over the sophistication of current A-share investors, and whether they can be relied upon to evaluate the worthiness of IPO firms.  Indeed, retail investors do play a large role in the trading of A-shares at the moment.  However, in view of the many problems associated with current IPO regulations, we believe investor protection arguments no longer justify retention of the status quo.  In fact, we believe it is likely that IPO reform will trigger large inflows of foreign capital as MSCI and other index providers increase their A-share allocation to full-weight (see Section 4 for details).  At the same time, this reform will aid in attracting the high-growth technology firms that investors, both domestic and foreign, are most interested in owning.  Financial education and investment literacy should improve quickly as domestic equity markets become more integrated with global markets.

*5.2 Reflections on Future Research Directions*

Based on our survey of current research, we have identified three under-studied areas.  While these fall under the general rubric of entrepreneurial finance, they seem particularly relevant to the current state of affairs in China.

First, relatively little work has been done on the interaction between financing choices (particularly public market financing) and innovations.  Given that much of industrial innovations is not accomplished by organizations other than small firms backed by venture capital (Kortum and Lerner 2000), it is important to understand the role that public markets and arms-length financiers play in facilitating innovation.  For example, Atanassov, Nanda, and Seru (2007) find that firms with arm's length financing have more patents and more novel patents. What drives this result?  Is this finding related to the distinction between exploitative versus exploratory innovation, as discussed in Ferreira, Manso, and Silva (2012)?  Does it reflect selection or treatment effects?  Some progress is being made in this area (e.g., Bernstein 2015; Acharya and Xu 2017), but more research is clearly needed.

Second, the role of government and interventions is understudied.  Researchers have begun to examine the ways in which policymakers can catalyze the growth of venture capital and encourage innovation (Irwin and Klenow 1996; Wallsten 2000).[27] Government interventions can help coordinate innovative effort which exhibits complementarity. At the same time, they can alter agents' incentives to become an entrepreneur and innovate, as well as the informational environment in general (Cong, Grenadier, and Hu 2018); Brunnermeier, Sockin, and Xiong 2017).

Given the prominent role of the government in the Chinese economy, it is all the more important to understand the benefits and unintended consequences of government interventions. As we reported in Section 3, government entities and state-owned-enterprises invest extensively in Chinese startups. Government initiatives such as the Innofund have also played an active role in financing R&D activities among small and medium technology firms. Yet, with few exceptions (e.g., Guo et al. 2016; Wang et al. 2017), prior studies have provided little evidence on the usefulness of such programs in improving innovation outputs. Of course, government-led programs also need to consider non-financial policy objectives, such as income inequality, market stability, job creation, or development of strategic sectors. Indeed, interventions designed to serve other objectives may have unintended consequences on the innovation process. Clearly, more research is needed to understand both the benefits and costs of government interventions, whether they are directly targeted at innovations or not.

Third, studies on the impact of new forms of financing innovation and startups are emerging. P2P lending, crowdfunding, and blockchain-based initial coin offerings are notable examples. How are projects evaluated in this environment? Is information better aggregated for efficient investment or termination of innovative projects? Do crypto-tokens serve special functions that encourage open-sourced, decentralized innovations? Mollick and Nanda (2015), Cong and Xiao (2017), and Cong, Li, and Wang (2018a,b) are early discussions on this front. In the Chinese setting, many of the new e-commerce and FinTech firms may provide the data needed to conduct such studies (e.g., Hau et.al. 2018). We view this as a rich area for future research.

---

[27] Lerner (2009) provides a review of the key government programs and their evaluations; Howell (2017) is a more recent analysis.

In sum, current research on financing innovations in China is quite limited.  Our paper aims to provide a modest contribution in this area by bringing together a wide set of related literature in economics and finance.  Given the government's active role in encouraging entrepreneurship and regulating the financial market, China offers an attractive context in which to advance our understanding on many of the aforementioned issues.

# Figure 1. Funding Sources for companies over their life cycle

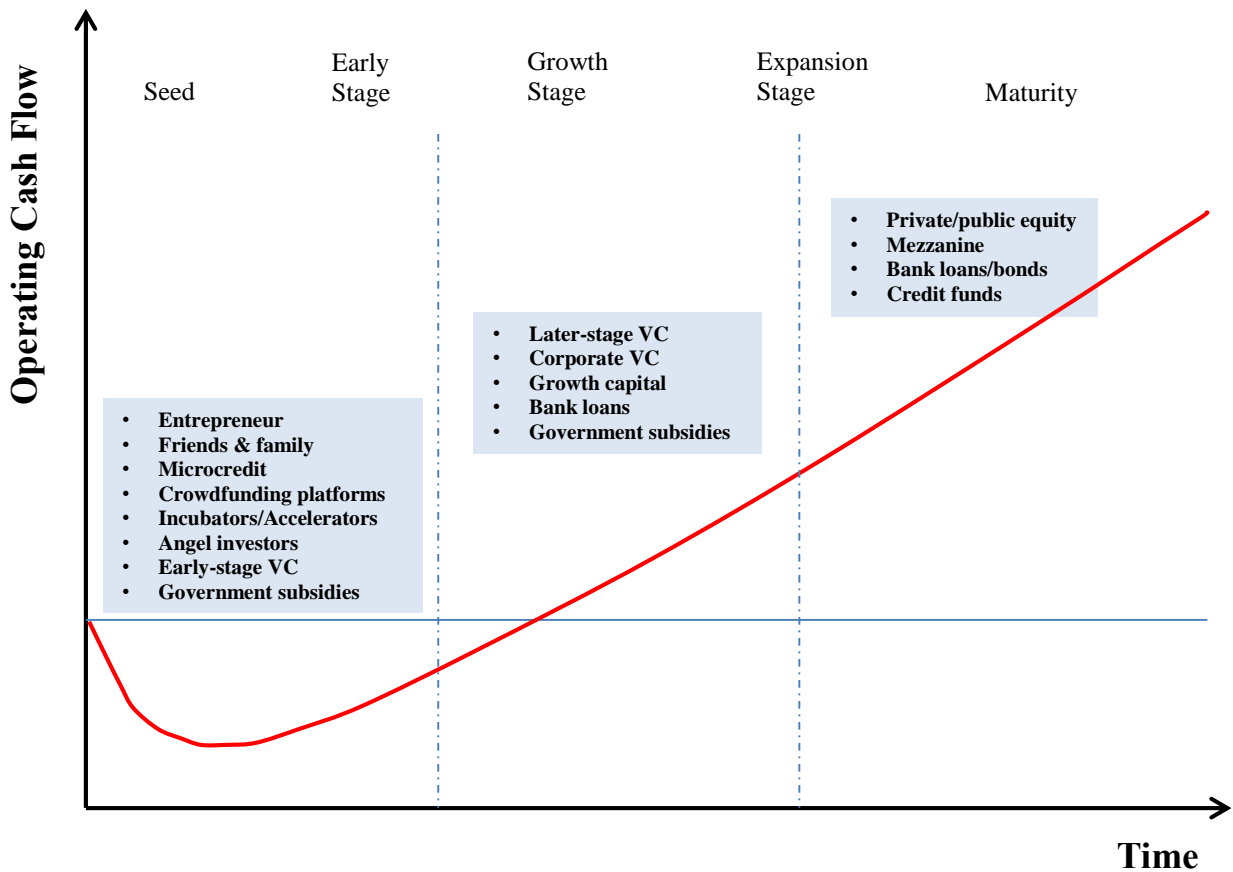

*Source: Adapted from Casanova, Cornelius, and Dutta (2018)*

**Chinese initiatives addressing companies' funding needs**

- Innofund – Financing for R&D of small to medium technology enterprises (SMTEs)

- VCPE - Venture Capital and Private Equity Funding

- CVC - Corporate investment, either directly, as a Strategic Partner; or indirectly, through a Venture Affiliate (e.g., Tencent and Alibaba)

- GVC - Government-led PEVC funds

- SOE - Direct Investment by State-owned Entities

- NEEQ – Listing on the National Equities Exchange and Quotations

**Table 1. Venture Capital and Private Equity (VCPE) Investment in Chinese Startup Firms**

This table reports the number of VCPE investment events in Chinese startup firms during the 2006 to 2017 period, and the dollar value of these investments. This information is extracted from PEdata. In this database, each bilateral investor-investee transaction is defined as a separate investment event. All Chinese startup firms in the database are included. We define Chinese firms as those whose headquarters are in China; similarly, Chinese (Foreign) VCPE firms are those headquartered in China (overseas). All monetary amounts in the table are in billions of U.S. dollars. Values originally reported in RMB or other foreign currencies were converted to US dollars using exchange rates at the end of each year. Investment values are reported as of each financing-round. When several funds participated in the same round but PEdata did not parse out the amount for each fund, we assume the funds contributed equally. We report total annual investments, as well as the cumulative amount invested over the entire 12-year period.

| | Chinese VC in Chinese Firms (Billion USD) | | | Foreign VC in Chinese Firms (Billion USD) | | | Total VC in Chinese Firms (Billion USD) | | |
|---|---|---|---|---|---|---|---|---|---|
| Year | N of Events | Annual | Cumulative | N of Events | Annual | Cumulative | N of Events | Annual | Cumulative |
| 2006 | 452 | 6.95 | 6.95 | 254 | 12.10 | 12.10 | 706 | 19.05 | 19.05 |
| 2007 | 862 | 9.23 | 16.17 | 292 | 7.45 | 19.55 | 1154 | 16.68 | 35.72 |
| 2008 | 1008 | 10.11 | 26.29 | 262 | 6.69 | 26.24 | 1270 | 16.80 | 52.52 |
| 2009 | 1217 | 18.85 | 45.13 | 171 | 3.08 | 29.31 | 1388 | 21.92 | 74.45 |
| 2010 | 2754 | 28.95 | 74.09 | 257 | 7.68 | 36.99 | 3011 | 36.63 | 111.08 |
| 2011 | 4273 | 47.18 | 121.27 | 328 | 8.74 | 45.73 | 4601 | 55.92 | 167.00 |
| 2012 | 3318 | 38.60 | 159.87 | 252 | 6.45 | 52.18 | 3570 | 45.05 | 212.05 |
| 2013 | 3004 | 44.59 | 204.45 | 220 | 4.52 | 56.70 | 3224 | 49.10 | 261.15 |
| 2014 | 4608 | 61.63 | 266.08 | 443 | 17.10 | 73.80 | 5051 | 78.73 | 339.88 |
| 2015 | 10373 | 115.85 | 381.93 | 436 | 10.84 | 84.64 | 10809 | 126.69 | 466.57 |
| 2016 | 9688 | 112.43 | 494.36 | 334 | 6.54 | 91.17 | 10022 | 118.96 | 585.53 |
| 2017 | 10241 | 144.65 | 639.01 | 385 | 7.05 | 98.22 | 10626 | 151.69 | 737.23 |
| Total | 51798 | 639.01 | | 3634 | 98.22 | | 55432 | 737.23 | |

**Table 2. Top Twenty Private Tech Firms ("Unicorns") in China**

This table lists the top 20-privately-owned technology firms ("Unicorns") in China as of the end of 2017, and their estimated valuations in billions of US dollars. The list is derived from China's Ministry of Science and Technology report titled "2017 China Unicorn Enterprise Development Report." For each firm, we also list the year of founding (Year), its industry section (Sector), whether it received funding from venture capital or private equity funds (VCPE) funding, whether it has a state-owned enterprise as an investor (SOE), and if so, the type of SOE. The funding information is collected from the PEdata database. Specifically, CIC denotes the China Investment Corporation (China's sovereign wealth fund); Corp denotes a government owned corporation, Pension denotes a pension fund, Local denotes a local government fund; and GVC denotes a government venture capital or private equity fund.

**Table 2.  Top Twenty Private Tech Firms ("Unicorns") in China  (continued)**

| No. | Company | Est Value (US$B) | Year | Sector | VCPE | SOE | Type of SOE |
|---|---|---|---|---|---|---|---|
| 1 | Ant Financial | 75.0 | 2014 | Internet Finance | Yes | Yes | CIC; Corp; Pension |
| 2 | Didi Chuxing | 56.0 | 2012 | Internet Service (Transportation) | Yes | Yes | CIC; Corp |
| 3 | Xiaomi | 46.0 | 2010 | Hardware | Yes | No | |
| 4 | Alibaba Cloud | 39.0 | 2009 | Internet Service (Cloud) | Yes | No | |
| 5 | Meituan-Dianping | 30.0 | 2010 | E-commerce | Yes | Yes | CIC; Corp |
| 6 | Ningde Shidai | 20.0 | 2011 | New energy car | Yes | Yes | Local |
| 7 | Jinri Toutiao | 20.0 | 2012 | New media | Yes | Yes | Corp |
| 8 | Cainiao Network | 20.0 | 2013 | Logistics/Transp | Yes | No | |
| 9 | Lufax | 18.5 | 2011 | Internet Finance | Yes | Yes | Corp |
| 10 | Jiedaibao | 10.8 | 2014 | Internet Finance | Yes | No | |
| 11 | Weizhong Bank | 9.2 | 2015 | Internet Finance | Yes | No | |
| 12 | Pingan Health | 8.8 | 2016 | Internet Finance (Health Insur) | Yes | No | |
| 13 | Koubei | 8.0 | 2015 | E-commerce | Yes | Yes | CIC |
| 14 | Jinrong Yizhangtong | 8.0 | 2015 | Internet Finance | Yes | No | |
| 15 | JD Finance | 7.7 | 2013 | Internet Finance | Yes | Yes | Corp |
| 16 | Ele.me | 5.5 | 2008 | E-commerce | Yes | Yes | Corp |
| 17 | Pin An Good Doctor | 5.4 | 2014 | Healthcare | Yes | No | |
| 18 | Weima Motor | 5.0 | 2011 | New energy car | Yes | Yes | GVC |
| 19 | Lianying Health | 5.0 | 2011 | Healthcare | Yes | Yes | GVC; Corp |
| 20 | Weilai Motor | 5.0 | 2014 | New energy car | Yes | Yes | Corp |
| | Total | 402.9 | | | | | |

**Table 3. Investments made in Chinese Startups by Tencent and Alibaba**

This table reports the value of investments made by Tencent and Alibaba in Chinese startups during the 2008 to 2018 period. Table values represent the number of financing events made each year and the value of these investments in US$ billions. This information is extracted from PEdata. Panel A reports only investments made through their VCPE funds; Panel B also includes strategic investments made through other corporate entities within the group. The Tencent group invests through four corporate entities and three VCPE funds. The most important vehicle is Tencent Capital, which accounts for more than 90% of the group's investments in startups. The Alibaba group invests through 6 firms and 2 VCPE funds. The most important vehicle is Ali Capital, which accounts for more than 90% of the investments made by the Ali group.

| Panel A: VC/PE Only | Tencent | | Alibaba | |
|---|---|---|---|---|
| Year | N of Events | Total Value | N of Events | Total Value |
| 2008 | 0 | 0.000 | 2 | 0.015 |
| 2009 | 1 | 0.003 | 3 | 0.009 |
| 2010 | 3 | 0.026 | 3 | 0.009 |
| 2011 | 20 | 0.460 | 6 | 0.081 |
| 2012 | 22 | 0.143 | 5 | 0.031 |
| 2013 | 20 | 0.150 | 21 | 2.738 |
| 2014 | 54 | 3.000 | 38 | 3.113 |
| 2015 | 75 | 3.345 | 43 | 5.838 |
| 2016 | 55 | 3.348 | 21 | 0.934 |
| 2017 | 88 | 8.138 | 20 | 2.952 |
| Total | 338 | 18.614 | 162 | 15.720 |

| Panel B: All Group Entities | Tencent | | Alibaba | |
|---|---|---|---|---|
| Year | N of Events | Total Value | N of Events | Total Value |
| 2008 | 2 | 0.014 | 2 | 0.015 |
| 2009 | 2 | 0.003 | 5 | 0.009 |
| 2010 | 6 | 0.036 | 5 | 0.009 |
| 2011 | 22 | 0.460 | 8 | 0.081 |
| 2012 | 26 | 0.212 | 7 | 0.031 |
| 2013 | 30 | 0.612 | 25 | 2.687 |
| 2014 | 82 | 3.768 | 49 | 3.530 |
| 2015 | 115 | 3.637 | 66 | 6.523 |
| 2016 | 90 | 3.612 | 48 | 2.342 |
| 2017 | 118 | 10.782 | 32 | 8.462 |
| Total | 493 | 23.137 | 247 | 23.689 |

**Table 4. Venture Capital and Private Equity (VCPE) Exit Events**

This table presents the means of exit for venture capital and private equity (VCPE) investors in Chinese startup firms. The data is extracted from PEdata. Each investor-investee transaction is a separate event in this database. An exit event is defined as the unwinding of a transaction previously entered in the database as an investment event. Panel A reports a breakdown of all exit events by category, where IPO denotes initial public offerings, Repurchase denotes shares reacquired by the investee, M&A is merger and acquisition, and Equity Transfer refers sales from one investee to another. Panel B reports a breakdown of IPO exits by stock exchange. In this panel, SHSE is the Shanghai Stock Exchange and SZSE is the Shenzhen Stock Exchange excluding the ChiNext Board. Although the ChiNext Board is officially part of the SZSE, it is separately reported here because listing requirements are lower and it is an important exit market for venture investors. US IPOs include listings in the NYSE, AMEX, and the NASDAQ. Hong Kong IPOs only include listings on the main board as the Growth Enterprise Market (GEM) is extremely illiquid. Note that the total number of IPOs in Panel B does not equal the total in Panel A because some exit events occur in other countries/regions, such as London or Singapore.

| Panel A: Means of Exit | | | | |
|---|---|---|---|---|
| Year | N of Events | IPO | Repurchase | M&A | Equity Transfer |
| 2006 | 138 | 69 | 1 | 50 | 18 |
| 2007 | 303 | 204 | 6 | 64 | 29 |
| 2008 | 138 | 58 | 7 | 44 | 29 |
| 2009 | 296 | 160 | 6 | 85 | 45 |
| 2010 | 671 | 488 | 22 | 101 | 60 |
| 2011 | 651 | 441 | 24 | 115 | 71 |
| 2012 | 613 | 274 | 62 | 163 | 114 |
| 2013 | 740 | 72 | 166 | 328 | 174 |
| 2014 | 1,534 | 388 | 142 | 665 | 339 |
| 2015 | 1,582 | 595 | 150 | 472 | 365 |
| 2016 | 1,318 | 566 | 75 | 336 | 341 |
| 2017 | 2,255 | 1,100 | 86 | 537 | 532 |
| Total | 10,239 | 4,415 | 747 | 2,960 | 2,117 |

**Table 4.  Venture Capital and Private Equity Exit Events (continued)**

| Year | Total | SHSE | SZSE | ChiNext | NEEQ | U.S. | Hong Kong |
|------|-------|------|------|---------|------|------|-----------|
| Panel B: IPO Exits by Exchange | | | | | | | |
| 2006 | 50 | 0 | 15 | 0 | 0 | 19 | 16 |
| 2007 | 178 | 8 | 49 | 0 | 1 | 77 | 43 |
| 2008 | 55 | 0 | 31 | 0 | 0 | 3 | 21 |
| 2009 | 155 | 0 | 40 | 50 | 0 | 19 | 46 |
| 2010 | 474 | 29 | 139 | 133 | 0 | 107 | 66 |
| 2011 | 437 | 43 | 142 | 150 | 0 | 64 | 38 |
| 2012 | 260 | 36 | 74 | 122 | 1 | 10 | 17 |
| 2013 | 71 | 0 | 1 | 0 | 3 | 24 | 43 |
| 2014 | 357 | 77 | 38 | 83 | 2 | 78 | 79 |
| 2015 | 518 | 181 | 83 | 178 | 0 | 15 | 61 |
| 2016 | 526 | 203 | 103 | 153 | 0 | 15 | 52 |
| 2017 | 926 | 412 | 143 | 328 | 0 | 32 | 11 |
| Total | 4,007 | 989 | 858 | 1,197 | 7 | 463 | 493 |

**Table 5. Initial Public Offering (IPO) and Reverse Merger (RM) Transactions**

This table presents the number and dollar value of Chinese firms that undertook an initial public offering (IPO) or a reverse merger (RM) transaction in the years 2007 to 2017. All monetary amounts are in billions of U.S. dollars. We include all firms headquartered in China that went public through an IPO in China Mainland, Hong Kong, or the U.S. For each IPO, we report the total funds raised, as well as the implied market valuation for the firm as a whole based on the IPO issue price. We also include firms that accessed the China Mainland market using a RM transaction as per Lee, Qu, and Shen (2018a). For each RM, we report the market value at the announcement of the RM transaction, as well as the current market value of the firm based on closing prices as of the end of March, 2018. The IPOs in mainland China include the IPOs in Shanghai and Shenzhen Stock Exchange. The data source is CSMAR. The Chinese firms IPOs in Hong Kong include the IPOs in Hong Kong Main Board. The data source is WIND and CSMAR. The Chinese firms IPOs in U.S. include the IPOs in AMEX, NYSE and NASDAQ. The data source is Compustat and SDC Plantinum.

| Year | IPO in China Mainland | | | IPO in Hong Kong | | | IPO in U.S. | | | RM in China Mainland | | |
|------|-----|-----------|---------|-----|-----------|---------|-----|-----------|---------|-----|-----------|------------|
| | N | Amt Raised | Mkt Val | N | Amt Raised | Mkt Val | N | Amt Raised | Mkt Val | N | Initial MV | Current MV |
| 2007 | 123 | 11.07 | 180.79 | 39 | 3.25 | 14.93 | 17 | 3.65 | 15.56 | 8 | 31.91 | 36.02 |
| 2008 | 76 | 2.38 | 14.79 | 17 | 0.87 | 5.47 | 4 | 0.25 | 1.27 | 11 | 11.34 | 24.49 |
| 2009 | 99 | 4.29 | 19.23 | 42 | 3.23 | 17.85 | 6 | 1.42 | 5.49 | 16 | 19.25 | 47.48 |
| 2010 | 347 | 10.85 | 66.68 | 63 | 3.97 | 28.39 | 23 | 2.68 | 13.50 | 12 | 13.69 | 27.85 |
| 2011 | 281 | 6.31 | 33.47 | 44 | 1.81 | 10.75 | 8 | 1.62 | 9.61 | 11 | 18.11 | 33.13 |
| 2012 | 154 | 2.28 | 12.86 | 33 | 1.37 | 8.45 | 3 | 0.16 | 1.02 | 10 | 8.51 | 18.42 |
| 2013 | 0 | 0.00 | 0.00 | 54 | 2.54 | 12.92 | 6 | 0.48 | 3.26 | 19 | 34.72 | 50.92 |
| 2014 | 125 | 1.45 | 10.34 | 67 | 2.95 | 15.37 | 10 | 24.81 | 203.07 | 23 | 39.54 | 68.51 |
| 2015 | 219 | 3.48 | 17.97 | 62 | 4.21 | 21.12 | 0 | 0.00 | 0.00 | 23 | 89.52 | 85.75 |
| 2016 | 227 | 3.29 | 20.97 | 41 | 2.75 | 16.20 | 3 | 1.54 | 14.80 | | | |
| 2017 | 436 | 4.92 | 29.34 | 35 | 1.58 | 9.54 | 5 | 1.49 | 12.46 | | | |
| Total | 2087 | 50.32 | 406.44 | 497 | 28.54 | 161.00 | 85 | 38.10 | 280.03 | 133 | 266.60 | 392.57 |

**Table 6. Key Characteristics of Mainland IPOs versus Other New Listings**

This table presents descriptive characteristics for the Chinese firms that choose to IPO in a Mainland exchange (SHZE or SZSE), or in Hong Kong, or the U.S., as well as those that elect to access a Mainland exchange through a reverse merger (RM) transaction. Assets is the reported total assets immediately before public listing, expressed in US$ billion. ROA is return-on-asset, defined as (Net Profit + Financing Expenses + Income Tax Expenses) divided by total assets. MB for IPOs is the market-to-book ratio, defined as the implied market value of the firm as a whole based on the IPO issue price, divided by (Book Value of Equity + Deferred taxes + Investment Tax Credits received in the pre-listing year). MB for RMs is the market value at the end of the initial announcement period, divided by pre-listing book equity. Lev is book leverage, defined as total liabilities divided by total assets. Cash is cash holdings divided by total assets. Following Lee, Qu, and Shen (2018a, 2018b), our RM sample includes all "clean RM transactions," wherein control of the combined entity is transferred to the owners of the unlisted firm, and the entire exchange takes place in a single transaction. Panel A reports the mean for each variable; Panel B reports the median. The Diff column reports the difference (Mainland IPO minus the firms identified in the column heading). ***, **, and * denote statistical significance at the 1%, 5%, and 10% level for a difference-in-mean (Panel A) or a difference-in-median (Panel B) test between the Mainland IPO firms and the firms identified in each column heading.

| Panel A: Mean | | | | | | | |
|---|---|---|---|---|---|---|---|
| | Mainland IPO | IPO in Hong Kong | | IPO in U.S. | | Mainland RM | |
| | mean | mean | Diff | mean | Diff | mean | Diff |
| Assets | 0.600 | 0.992 | -0.392** | 0.365 | 0.235 | 0.725 | -0.125 |
| ROA | 0.159 | 0.160 | -0.001 | 0.060 | 0.099*** | 0.146 | 0.013* |
| MB | 1.148 | 1.278 | -0.130* | 74.206 | -73.06*** | 6.322 | -5.17*** |
| Lev | 0.447 | 0.614 | -0.168*** | 0.466 | -0.019 | 0.536 | -0.090*** |
| Cash | 0.204 | 0.178 | 0.026*** | 0.327 | -0.123*** | 0.173 | 0.031** |

| Panel B: Median | | | | | | | |
|---|---|---|---|---|---|---|---|
| | Mainland IPO | IPO in Hong Kong | | IPO in U.S. | | Mainland RM | |
| | median | median | Diff | median | Diff | median | Diff |
| Assets | 0.092 | 0.205 | -0.114*** | 0.096 | -0.004 | 0.416 | -0.324*** |
| ROA | 0.146 | 0.127 | 0.020 | 0.111 | 0.035*** | 0.113 | 0.034*** |
| MB | 0.876 | 0.705 | 0.171*** | 13.668 | -12.792*** | 2.851 | -1.974*** |
| Lev | 0.447 | 0.636 | -0.189*** | 0.440 | 0.007 | 0.561 | -0.115*** |
| Cash | 0.172 | 0.143 | 0.029*** | 0.298 | -0.127*** | 0.141 | 0.031*** |

**Table 7. NEEQ Market Overview**

This table presents an overview of the National Equities Exchange and Quotations (NEEQ) Market. Total market capitalization and total trading volume are in billions of U.S. dollar. In Panel A we provide summary statistics on market liquidity. Total trading volume is further subdivided into trades that involve a market maker, and trades that were consummated through bilateral negotiation between the parties. In Panel B, for comparison, we also present market liquidity statistics for the A-Share markets (Shanghai and Shenzhen combined). In Panel C we report a distribution of the number of investors for the NEEQ firms, as of the end of 2017.

| Panel A: NEEQ Market Liquidity | | | | | | | |
|---|---|---|---|---|---|---|---|
| Year | No. of firm | Total Mkt Cap | Avg Mkt Cap | Annual Turnover | Total Trading Volume | Volume: Market Makers | Volume: Bilateral Negotiations |
| 2013 | 356 | 8.51 | 0.02 | 0.04 | 0.13 | 0.00 | 0.13 |
| 2014 | 1,572 | 70.64 | 0.04 | 0.20 | 2.01 | 0.33 | 1.68 |
| 2015 | 5,129 | 378.22 | 0.07 | 0.54 | 29.39 | 17.00 | 12.33 |
| 2016 | 10,163 | 623.97 | 0.06 | 0.21 | 29.42 | 14.61 | 14.80 |
| 2017 | 11,630 | 760.07 | 0.07 | 0.13 | 34.95 | 12.07 | 22.87 |

| Panel B: A-Share Market Liquidity | | | | | |
|---|---|---|---|---|---|
| Year | No. of firm | Total Mkt Cap | Avg Mkt Cap | Annual Turnover | Total Trading Volume |
| 2013 | 2,470 | 3670.37 | 1.49 | 2.34 | 7,139.20 |
| 2014 | 2,592 | 5735.99 | 2.21 | 2.34 | 11,342.97 |
| 2015 | 2,811 | 8224.77 | 2.93 | 6.06 | 39,076.45 |
| 2016 | 3,033 | 7795.66 | 2.57 | 3.23 | 19,364.67 |
| 2017 | 3,465 | 8695.24 | 2.51 | 2.40 | 16,492.62 |

| Panel C: Number of Investors for NEEQ Firms | | |
|---|---|---|
| No. of investors | No. of firms | Percentage |
| 2 | 742 | 6.38% |
| 3-10 | 4,454 | 38.30% |
| 11-50 | 4,529 | 38.94% |
| 51-100 | 953 | 8.19% |
| 101-200 | 551 | 4.74% |
| 200 and above | 401 | 3.45% |
| Total | 11,630 | 100.00% |